# A new helical InSeI polymorph: crystal structure and polarized Raman spectroscopy study

Lucía Olano-Vegas,[a,b] Davide Spirito,[c] Evgeny Modin,[a] Pavlo Solokha,[d] Sergio Marras,[e] Marco Gobbi,[f,g] Fèlix Casanova,[a,g] Serena De Negri,[d] Luis E. Hueso,[a,g]* and Beatriz Martín-García,[a,g]*

[a.] *CIC nanoGUNE BRTA, 20018 Donostia-San Sebastián, Basque Country, Spain.*
[b.] *Departamento de Polímeros y Materiales Avanzados: Física, Química y Tecnología, University of the Basque Country (UPV/EHU), 20018, Donostia-San Sebastian, Spain*
[c.] *BCMaterials - Basque Center for Materials, Applications, and Nanostructures, UPV/EHU Science Park, Leioa 48940, Spain.*
[d.] *Dipartimento di Chimica e Chimica Industriale, Università degli Studi di Genova, Genova 16146, Italy.*
[e.] *Istituto Italiano di Tecnologia - Materials Characterization Facility, Genova 16163, Italy.*
[f.] *Materials Physics Center CSIC-UPV/EHU, 20018 Donostia-San Sebastián, Spain.*
[g.] *IKERBASQUE, Basque Foundation for Science, 48009 Bilbao, Spain.*
*Email: l.hueso@nanogune.eu; b.martingarcia@nanogune.eu

**Abstract:** Tetragonal InSeI is an interesting low-dimensional metal chalcohalide due to its composition and anisotropic crystal structure composed of helical chains, which give rise to optoelectronic properties with potential application in photodetectors, optical thermometers, and spintronic devices. However, experimental works lack on the study of its anisotropic or chiral behavior. Here we present the crystal structure of an unreported InSeI polymorph and study its lattice dynamics in bulk crystals and exfoliated nanowires by polarized Raman spectroscopy for two non-equivalent crystallographic planes. We determine the orientation of the helical chains and distinguish between crystallographic planes by linearly polarized measurements, evaluating the angle-dependent intensity of the modes, which allows assigning each mode to its representation. Circularly polarized Raman measurements do not reveal chiral phonons, despite the helical chains and anisotropic crystal structure. These results offer insight into the crystal structure of InSeI, which is fundamental for the fabrication of orientation-dependent optoelectronic and spintronic devices.

## Introduction

Metal chalcohalides have recently attracted the attention of researchers due to the diversity of their crystal structures and to their optical, electrical, magnetic, thermal and catalytic properties, which give these materials a broad range of potential applications. These include photovoltaics, thermoelectrics, photocatalysis, radiation detection and spintronics.[1–8] The composition and crystal structure of this family of materials has been known since at least the 1960s, but they have only been recently identified as materials with potential for on-demand properties. These compounds share the general formula MEX, where M is a metal cation (Ag, Cu, Sn, Pb, Cr, In, Ga, Sb, As, Bi and others), and E and X are chalcogen (O, S, Se, Te) and halogen (Cl, Br, I) anions, respectively.[1–10] InSeI stands out among metal chalcohalides due to its unique tetragonal crystal structure, reported experimentally with the $I4_1/a$ space group.[11] It grows as a 1D van der Waals (vdW) material formed by helical chains and can be mechanically exfoliated into nanowires (NWs).[9,11–13] These helical chains give InSeI its anisotropic structure, making it a potential candidate to show crystallographic orientation dependent properties similar to those reported by other anisotropic 2D or 1D materials such as black phosphorous, ReS$_2$, group IV monochalcogenides,[14] ZrTe$_5$,[15] or tellurium[16,17]. Among the many promising properties of InSeI, its ferroelectricity is interesting for the fabrication of non-volatile memories.[18] Its high optical absorption in the UV range (<400 nm) and tunable bandgap (from bulk to single NW) make it suitable for photodetectors and photovoltaic devices.[13,19] Its thermochromic behavior enables its use in optical thermometers.[20] Lastly, it has been predicted to show helical-induced and antisymmetric spin-orbit coupling[21,22] and chirality-induced spin splitting,[23] which open avenues for its integration in spintronic devices. While this impressive combination of properties has sparked interest, the study of InSeI is still in its early stages. The existing body of research is mainly made up of theoretical studies of electronic states, chirality-induced spin selectivity, and optoelectronic properties.[19,21–23] While some experimental works do exist, these focus on optoelectronic properties without paying attention to either anisotropic or chiral effects.[13,20] The correlation of these directional properties with vibrational modes in the crystal requires a specific insight which can be provided by Raman spectroscopy.[24–26] Current reports on InSeI have been limited to a specific polymorph (*I*4$_1$/*a* space group),[11,20] however a rich knowledge can be obtained by the study of polarized spectra exploring various crystal orientations. Here, we investigate the anisotropic lattice dynamics with incidence in two non-equivalent lattice planes by linear and circular polarized Raman spectroscopy, both in bulk crystals and exfoliated NWs. Supported by X-ray diffraction (XRD) and scanning transmission electron microscopy (STEM), we identify a new InSeI polymorph. Results from angle-resolved linearly polarized Raman spectroscopy show a correlation between the angular dependence of some Raman modes' intensity and the orientation of the helical I-In-Se chains. Moreover, the different angular dependence of the vibrational modes' intensity, supported by Raman tensor analysis, allows identifying two types of crystallographic planes: those parallel and perpendicular to the helical chains. Circularly polarized Raman measurements reveal that the presence of helical chains and anisotropy in the crystal structure do not lead to observable chiral phonons. This knowledge allows a reliable and efficient identification of the crystal faces and orientation, key for the development and design of novel optoelectronic and spintronic devices which make use of anisotropic properties and low-dimensional effects.

**Experimental**

**Materials:** InSeI bulk fibrous-like wires used in this work were purchased from 2D Semiconductors and as indicated by the supplier they were synthesized through physico-chemical techniques used in crystal growth.

**Optical microscope images** of the crystals were taken with a Nikon Eclipse LV150A, and a Nikon DS-Ri2 optical microscope, using different objectives from 5× to 50× in bright field.

**X-ray diffraction characterization and analysis:** Several single crystals were extracted from the InSeI bulk sample, cut with a razor and selected under an optical microscope to be tested for XRD. Crystals were glued on microloops, then mounted on the goniometer head of a three-circle Bruker D8 QUEST diffractometer equipped with a PHOTON III photon counting detector and using the graphite monochromatized Mo Kα radiation (λ = 0.71076 Å). A fast scan procedure, consisting of a quick 180° f-scan, was applied to screen crystals quality. The selected crystals were all very thin and with a fibrous morphology, affecting the quality of the recorded diffraction pattern, frequently showing low intensities, effects related to twinning, multiple domains, etc. A complete dataset (three f-scans and one w-scan) was recorded on the best candidate for structure solution. Intensity data were collected over the reciprocal space up to ~ 20° in θ (corresponding to a ~ 1.05 Å resolution) with exposures of 10 s per frame. The frames were integrated and corrected for absorption effects with the Bruker software package[27,28]. The structure was solved with the intrinsic phasing algorithm implemented in APEX5[29] and refined in the space group $P\bar{4}$. The final full-matrix least-squares isotropic refinement converged to somewhat high residuals (R1 = 15.90%), which is a consequence of the abovementioned difficulties in getting good quality crystals.

Powder XRD patterns were recorded on a Malvern-PANalytical 3rd generation Empyrean X-ray diffractometer, equipped with a 1.8kW Cu ceramic X-ray tube ($\lambda_{CuK\alpha}$ = 0.15406 nm) operating at 40 kV and 45 mA, W/Si elliptic focusing mirror, GaliPIX$^{3D}$ solid-state pixel area detector. The sample was sealed between two layers of 7 μm-thick Kapton® foil lined with vacuum grease. The diffraction patterns were collected at room temperature in transmission geometry and 1D mode using a transmission spinner sample stage (rotation speed = 2 rps).

**Mechanical exfoliation of the InSeI crystals and deterministic all-dry viscoelastic stamping:** We exfoliated the InSeI bulk crystals by using scotch tape (Nitto® blue tape, BT-150E-KL) and transferred the obtained NWs to polydimethylsiloxane (PDMS, Gelfilm from Gelpak®) stamp. From the PDMS stamp, we transferred the NWs using an all-dry viscoelastic transfer system connected to an optical microscope system[30] mounted inside a glovebox, which allows to place the NWs on top of the target substrates, such as SiO$_2$/Si or SiN grids.

**Electron microscopy characterization:** Scanning transmission electron microscopy was used to visualize the crystal structure of mechanically exfoliated InSeI NWs stamped on SiN TEM grids by deterministic all-dry viscoelastic stamping as explained before. A Titan 80-300 transmission electron microscope (Thermo Fisher, USA) operating at an accelerating voltage of 300 kV was used. Due to the beam sensitivity of InSeI, fast scanning with a low electron dose was employed to capture images of the atomic structure. The dwell time was set to 0.5 μs, and approximately 50 images were acquired in a series. The images were then precisely aligned to eliminate drift and summed to enhance the signal-to-noise ratio. All image processing was performed using Gatan Digital Micrograph 3.5 software. The simulation of the potential projection magnitude was carried out using the ReciPro package.

**Raman spectroscopy characterization:** Micro-Raman characterization was carried out in a Renishaw® inVia Qontor micro-Raman instrument. To minimize the air-exposure of the InSeI material during room-temperature measurements an N$_2$ flow was used. For low-temperature measurements, we used a 50× objective (Nikon, WD 11 mm, N.A. 0.60) and a Linkam® vacuum chamber ($10^{-6}$ hPa) cooled by liquid N$_2$, enabling material characterization in the range of 80 - 300 K. We collected the Raman spectra using 532 nm as excitation source and a diffraction grating of 2400 lines/mm. The laser incident power was kept < 2 mW to avoid damage to the crystals during the acquisitions. Angle-dependent linearly polarized measurements were carried out at room temperature rotating the samples in a piezo-motor (Thorlabs, ELL18/M rotation stage – PC controlled by ELLO® software) placed on the Raman instrument XYZ stage and using a half-wave plate (HWP) (Renishaw®) in the excitation path to rotate the polarization of the excitation laser from vertical (V) to horizontal (H), and a linear polarizer in the detection path in vertical (V) configuration. For the circularly polarized measurements, we combined a HWP in front of the laser exit changing the linear polarization from V to H with a quarter-wave plate (QWP) just before the Rayleigh filters in the excitation path (Renishaw®) and a rotating QWP (Thorlabs, AQWP10M-580) with a fixed linear polarizer (Thorlabs, LPVISA100-MP2) in the detection path. The fast axis of the QWP was placed at +45° (σ$^+$) and -45° (σ$^-$) view along the propagation direction as seen from the sample using a piezo-motor (Thorlabs, ELL14K rotation mount – PC controlled by ELLO® software). In this way, we access the different configurations: σ$^-$, σ$^+$, σ$^-$σ$^+$ (σ$^-$-incident and σ$^+$-scattered) and σ$^+$σ$^-$ (σ$^+$-incident and σ$^-$-scattered), being σ$^-$ (right - clockwise) and σ$^+$ (left - anticlockwise). Python and Julia codes were employed to analyze the data collected fitting the Raman spectra datasets with Lorentzian functions. We checked the reproducibility of our results by measuring at least 5 different bulk crystals and 5 bulky NWs in several regions. We carried out control experiments to check the set-up and measurement conditions.

**Results and discussion**

In this work, we studied the lattice dynamics of InSeI commercial bulk crystals and mechanically exfoliated NWs. An initial check of the material quality by X-ray powder diffraction evidenced a strong preferred orientation along the [001] direction, arising from the 1D nature of the title compound, but it revealed a clear mismatch with the structural model proposed in the literature[11] (*tI*96, space group *I*4$_1$/*a*, no. 88). Therefore, we exploited the single crystal XRD to elucidate the crystal structure. Overcoming the difficulties arising from the fibrous morphology of the InSeI crystals (**Fig. S1**), we obtained a reasonable structural model in the tetragonal $P\bar{4}$ space group (no. 81) with *a* = 18.929(7) and *c* = 24.058(10) Å (see **Tables S1** and **S2**). Despite the somewhat high residuals of the single crystal refinement, the calculated powder diffraction pattern for our new model matches much better with the experimental one (**Fig. 1**a) compared to the *tI*96 model (**Fig. S2**). The chemical reliability of our result is also strengthened by different observations: i) for the analogous InSI compound two polymorphs were proposed in the literature[12,31], the *tI*96 and a new tetragonal one with *a* = 18.65 and *c* = 23.24 Å, without further details on the crystal structure; ii) the GaSI was recently reported to crystallize in

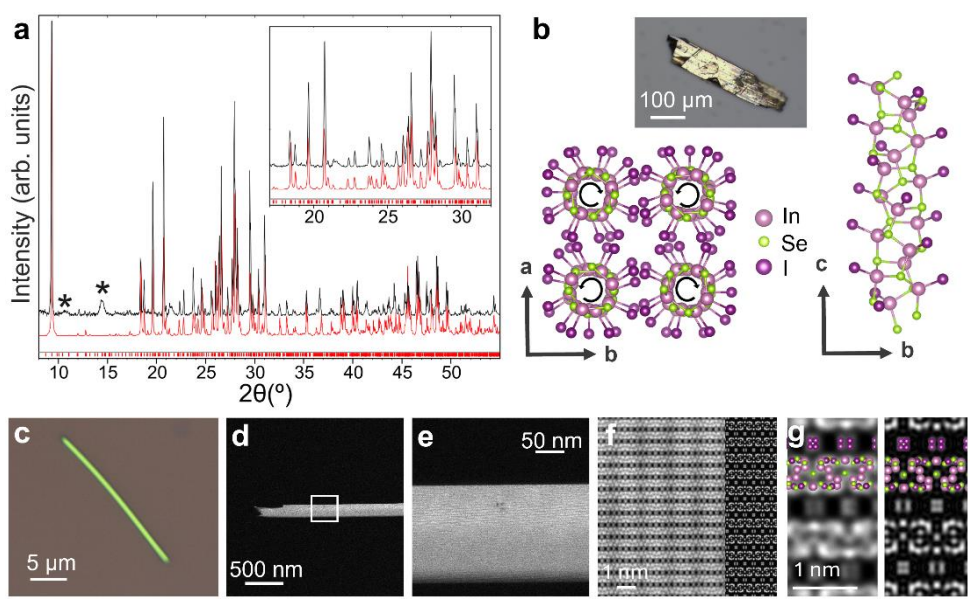

***Fig. 1.*** *Material characterization. a) Experimental X-ray powder diffraction pattern of InSeI collected in transmission mode, avoiding the preferred orientation effect (black spectrum) compared to the theoretical*

*pattern for the P$\bar{4}$ structural model proposed here (red spectrum). Peaks labelled with * are due to Kapton film. b) Optical image of a representative InSeI bulk crystal together with the visualization of the crystal structure showing an InSeI chain along the c-axis together with the picture of the chains down the c-axis. c) Optical image of a mechanically exfoliated nanowire. d-g) STEM image of an InSeI nanowire with the successive zooms to show the atoms arrangement (left) accompanied by the simulation (right) in (f) and (g) panels.*

the P$\bar{4}$ space group[32], with a doubled *c* parameter with respect to that obtained here for InSeI. All these facts confirm that we are dealing with a new InSeI polymorph. Similarly to the others,[9,11,13,20] this structure is composed of 1D helical chains comprised of covalently bonded In and Se, with I atoms bonded to the In ones and radially pointing out of the helix (Fig. 1b). The non-centrosymmetric unit cell contains four helices of opposed handedness arranged in a primitive and achiral way, with vdW forces holding them together. The principal difference with respect to InSeI (*I*4$_1$/*a*) stays in a slight distortion of helices causing the symmetry reduction and consequently a noticeable increase of the *c*-parameter. Thus, the 1D nature of the material is preserved and this allows its exfoliation in the form of NWs (Fig. 1c), with thickness ranging from 40 nm to 500 nm (**Fig. S3**). In the case of NWs, we determine their crystal orientation by STEM imaging (Fig. 1d-g). The experimental STEM images and the averaged image of the unit cell matched with the simulated images for the equivalent (100) and (010) crystallographic planes (**Fig. S4**). This confirms that the material exfoliates in NWs with [001] orientation in which the bulk crystals grow. Therefore, by using bulk crystals and NWs, we can observe two types of crystallographic planes: the equivalent (100) and (010), parallel to the I-In-Se helical chains, and (001), perpendicular to the chains and only accessible in the bulk crystals.

Once we identified the crystal structure and the different crystallographic planes in the InSeI bulk crystals and exfoliated NWs, we carried out Raman spectroscopy measurements (**Fig. 2**a). The Raman spectra collected at room temperature show 8 modes centered at 143±2 cm$^{-1}$, 157±2 cm$^{-1}$, 168±3 cm$^{-1}$, 183±3 cm$^{-1}$, 189±2 cm$^{-1}$, 200±3 cm$^{-1}$, 208±2 cm$^{-1}$ and 213±2 cm$^{-1}$ in the bulk crystals. Only the mode at 143 cm$^{-1}$ is well resolved in the case of the exfoliated NWs (**Fig. S5**), most probably due to a Raman signal thickness-dependent effect. Additionally, we do not observe variations in the number of Raman modes when we decrease the temperature down to 80 K, but only a systematic shift towards higher frequencies without discontinuities (**Fig. S6**). The frequency range of the modes, as well as the Raman signal shape observed in this InSeI polymorph, match well with the data reported for InSeI with space group *I*4$_1$/*a*.[20] This is not unexpected, since the crystal structure of both polymorphs is similar in terms of atoms arrangement and In-I and In-Se bonds distance. Therefore, we can assign the observed Raman modes to vibrations related to the In, Se and I atoms as predicted for the other polymorph. Being more specific, the sharp and strongest mode observed at 143 cm$^{-1}$ can be assigned to the breathing vibration of Se atoms with respect to the chain axis (Fig. 2b).[20] However, in accordance with group theory consideration applying for InSeI crystal's symmetry (point group S$_4$, Wyckoff position *4h* (x, y, z))[33–36], in this material more phonon active modes are expected than for the InSeI (point group C$_{4h}$, Wyckoff position *16f*).

To gain insight into the character of the Raman modes experimentally observed, we carried out a detailed polarized Raman spectroscopy study with different configurations and tensor analysis. We performed angle-resolved polarized Raman measurements using the set-up depicted in **Fig. 3**a (see more details in the Experimental section and control experiments in SI – **Fig. S7**) and evaluated the signal intensity of the well-resolved modes experimentally observed.

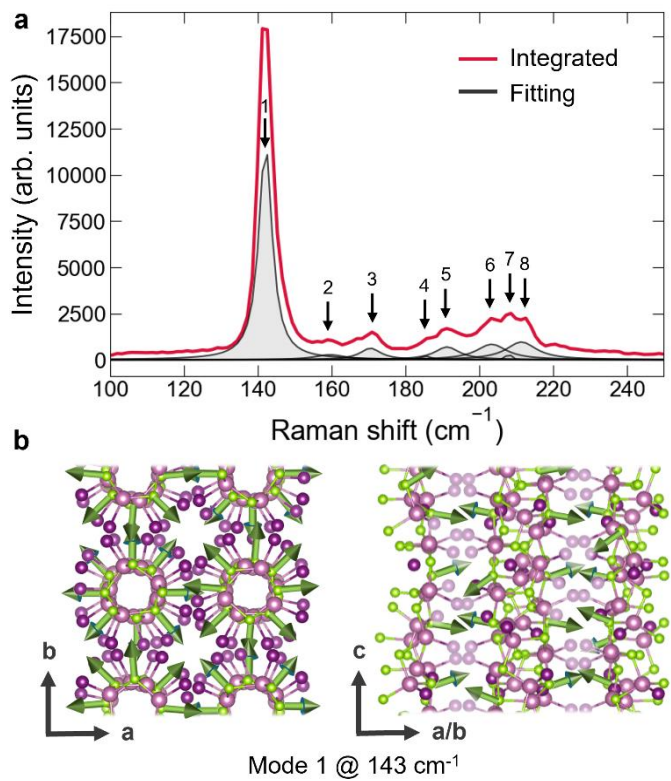

***Fig. 2.*** *a) Integrated Representative Raman spectrum of an InSeI crystal showing 8 active Raman modes fitted to a Lorentzian function. The spectrum was obtained as the sum of the spectra collected rotating the crystal with respect to the incident laser polarization over a range from 0º to 180º. The intensity values of the fitting have been reduced by 4% for the clarity of the figure. b) Sketches adapted from ref [20] of the Se atoms vibrations in the crystal structure related to the strongest Raman mode observed centered at 143 cm$^{-1}$. In the right panel, I atoms became transparent for better visualization of the arrows.*

Briefly, the sample is excited with a vertical (V) or horizontal (H) linearly polarized laser. A polarizer at the spectrometer is aligned
with the V direction, so that we have a "parallel" VV configuration or a "cross" HV configuration. The sample is mounted to access a specific crystallographic plane, and it is rotated at different azimuth angle (θ). In this way, we could correlate the measured Raman intensity as a function of θ with the crystal structure orientation. In the case of 1D NWs, in accordance with the STEM results, we collect the Raman signal on the equivalent (100) or (010) planes (Fig. 3b), where the helical chains are arranged parallel to these planes. We evaluate the Raman mode centered at 143 cm$^{-1}$ (mode 1), whose angle-dependent integrated intensity signal under the VV configuration leads to an ellipsoidal polar plot, whose maximum is aligned with the helical chains of the NW (Fig. 3c). This allows for a reliable crystal orientation determination. In contrast, under HV configuration, a four-lobe polar plot is observed, whose lobes are 45º rotated with respect to the lobes under VV configuration, not allowing a direct correlation of the polar plot pattern and the NW crystallographic orientation. To assign the symmetry of the modes, we compare the experimental angular pattern with results of Raman tensor analysis (see SI). We consider the position of the lobes and relative shift with respect to the helical chain direction in both VV and HV configurations. In the case of mode 1, at 143 cm$^{-1}$, since the ellipsoidal pattern in the VV polar plot is aligned with the helical chains, this excludes B and E representations, leading to an A mode. This is also in accordance with the four lobes in the HV polar tilted 45º vs VV polar plot, and results reported in InSeI (space group $I4_1/a$) using first-principles calculations, with $A_g$ assignation (A in the $S_4$ point group).[20] To evaluate the other active Raman modes in this material, we draw on the bulk crystals.

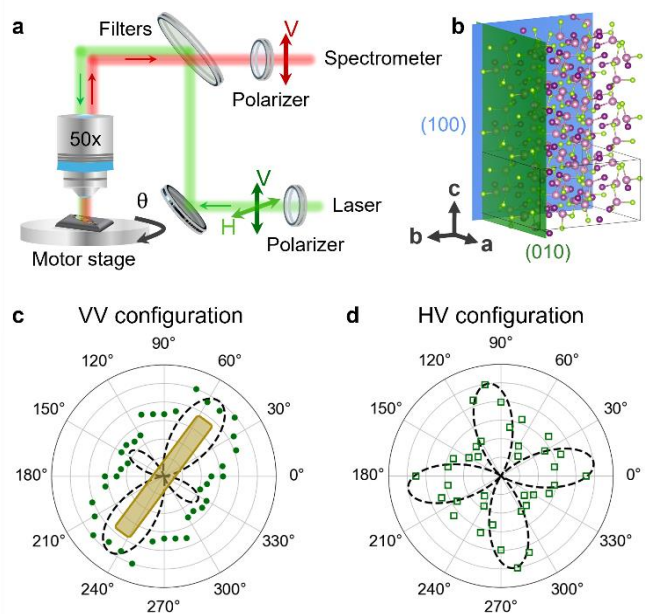

*Fig. 3. Linearly polarized Raman spectroscopy study in NWs. a) Schematic representation of the set-up used for the angle-resolved linearly polarized Raman measurements: parallel (VV) and cross (HV) configurations. The laser falls with vertical or horizontal linear polarization on the sample which rotates, θ, with the motor stage. The backscattered signal goes again through a polarizer with axis vertically allowing to detect only this component by the spectrometer. b) Representation of the (100) (blue) and (010) (green) crystallographic planes of the InSeI. The polar plots c) and d) show the normalized intensity of mode 1 as a function of the relative angle, θ, between the NW and the polarization direction of the laser, for the VV and HV configurations respectively. The black dashed lines show the result of Raman tensor analysis detailed in SI. The yellowish rectangle in panel c indicates the orientation of the NW.*

For bulk crystals, we are able to evaluate the angle-dependent intensities of six well-resolved modes centered at 143 cm$^{-1}$ (mode 1), 168 cm$^{-1}$ (mode 3), 183 cm$^{-1}$ (mode 4), 189 cm$^{-1}$ (mode 5), 200 cm$^{-1}$ (mode 6) and 208 cm$^{-1}$ (mode 7) in two different crystallographic planes (100)/(010) and (001), using the same set-up as for NWs (**Fig. 4**, and **Fig. S8** for data collected from the (001) plane). Focusing on the (100)/(010) planes, mode 1 centered at 143 cm$^{-1}$, the corresponding polar plots show the same trend as the one observed in NWs, and therefore, can be ascribed to an A mode. This assignation is further confirmed by the pattern of the polar plots collected on the (001) plane, with an isotropic behavior for VV configuration and a low signal in HV configuration. Remarkably, the patterns of the polar plots correlate well with the Se atoms vibrations displayed in Fig. 2, which are aligned with the chains in the (100)/(010) planes and are radial in the (001) plane. The modes 3 and 4, centered at 168 and 183 cm$^{-1}$ respectively, show polar plots in HV configuration in (100)/(010) planes with patterns similar to mode 1, indicating that they are A or B modes. This is further confirmed with the VV configuration showing a pattern which does not match with a B mode. Therefore, both modes 3 and 4 are A modes. Most interesting, as happened with mode 1, the maximum of the polar plot in VV configuration for mode 4 at the (100)/(010) planes is oriented with the helical chain axis. Mode 4 is zero in direction perpendicular to the helical chain, suggesting that the vibration is aligned with the chain. The mode 5, centered at 189 cm$^{-1}$, presents in the (100)/(010) plane an angle-dependence in HV configuration which is 45º rotated vs HV polar plot of mode 1, as happens for an E mode. This E character is proved by the VV and HV polar plots of the (001) plane, which are almost zero intensity. Looking at (100)/(010) VV and HV polar plots for modes 6 and 7, centered at 200 and 208 cm$^{-1}$, respectively, show a pattern which matches with A or B representations, being B the most plausible since the two lobes in VV configuration are perpendicular to helical chains at 0º. However, results in the (001) plane indicate VV and HV quadrupolar plots and 45º related, confirming B character for mode 6; while the isotropic VV polar plot and almost zero HV polar plot for mode 7 indicate an A mode. The representation designation for modes 6 and 7, as occurs with mode 1, is also in accordance with simulations reported for InSeI (space group $I4_1/a$), with $B_g$ and $A_g$ assignation for modes at similar frequency, respectively.[20] Therefore, the angle-dependent linearly polarized Raman spectroscopy measurements shown here can be used as tetragonal InSeI fingerprints to distinguish between the crystallographic planes parallel [(100)/(010) – equivalent planes] and perpendicular [(001)] to the helical chains, determining the crystallographic orientation of the bulk crystal and exfoliated NWs. Moreover, the Raman tensor theory analysis carried out to correlate the angular dependence of the different modes with the symmetry rules shows a good match with our experimental data as well as first-principles calculations[20] reported for the other polymorph. This corroborates that our experimental and theoretical methodology can be extended to the study of other tetragonal InSeI polymorphs,[11,20] the GaSI or GaSeI helical crystals,[9,32] or any crystal with point groups $C_4$, $S_4$ and $C_{4h}$, which have the same Raman tensors.

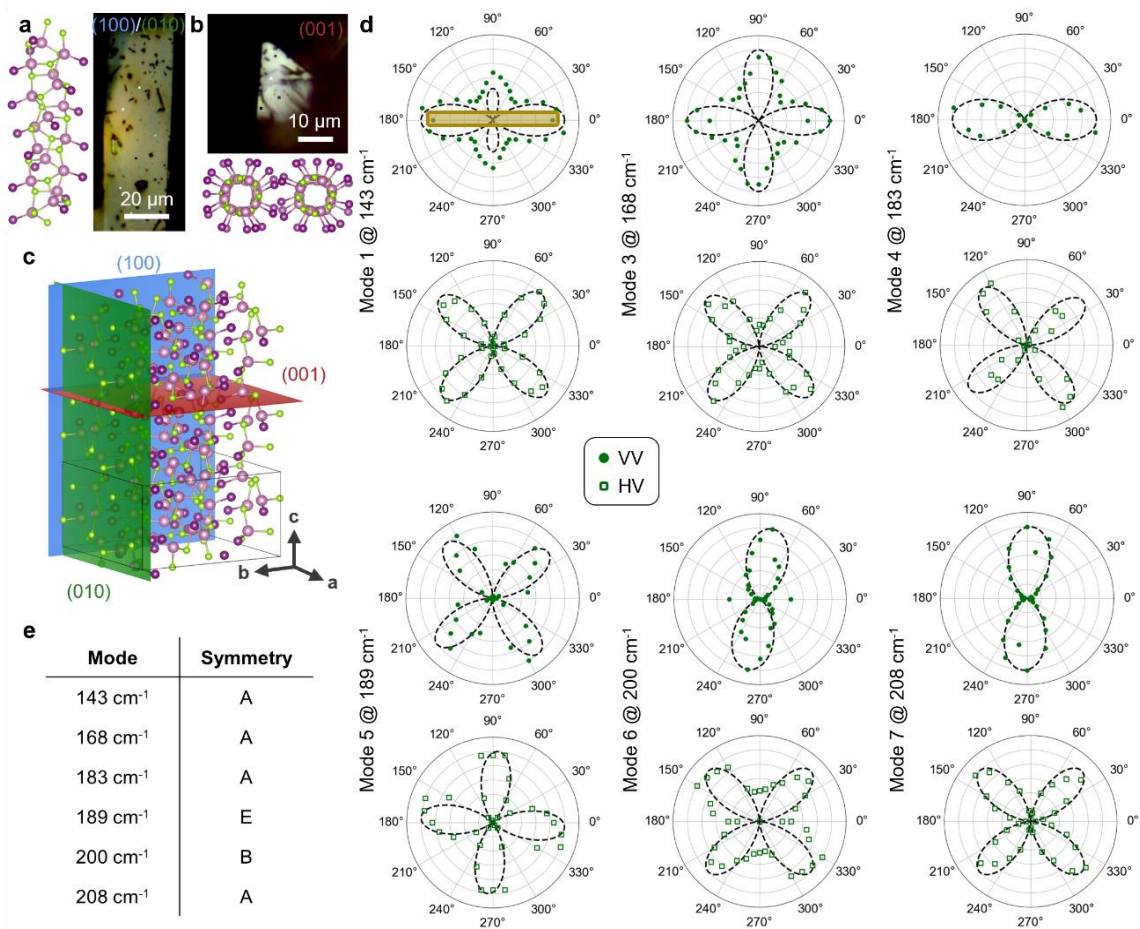

*Fig. 4. Linearly polarized Raman spectroscopy study in bulk crystals. a-b) Optical microscope image of the (100)/(010) and (001) planes of a crystal, respectively, with the corresponding visualization of the arrangement of the atoms. c) Representation of the (100) (blue), (010) (green) and the (001) (red) planes under study. d) Angle dependence of the normalized intensity of modes 143, 168, 183, 189, 200 and 208 cm$^{-1}$ under VV and HV configurations on the (100)/(010) plane. The experimental data is shown in full dots for the VV configuration and empty squares for the HV configuration, while the black dashed lines show the result of Raman tensor analysis detailed in SI. e) Assigned symmetries of the modes as determined by comparison with patterns obtained from Raman tensor analysis.*

Considering that the InSeI presents helical chains and anisotropy in its crystal structure, we study the response for the two crystal faces by carrying out circularly polarized Raman spectroscopy measurements (**Fig. 5** for bulk crystals, see **Fig. S9** for NWs). We explore two configurations to collect the Raman signal: (i) with circularly polarized incident light while collecting the unpolarized signal ($\sigma^-$/- vs $\sigma^+$/-); and (ii) by helicity-resolved Raman spectroscopy, using opposite circularly polarized incident and collected light ($\sigma^-\sigma^+$ vs $\sigma^+\sigma^-$, cross-helicity) (Fig. 5a, description in Experimental Section in SI). When comparing the Raman spectra collected with the configurations $\sigma^-$/- vs $\sigma^+$/- or in cross-helicity, even at low-temperatures with narrower peaks (**Fig. S10**), we do not observe any differences in the relative intensity or Raman shift or splitting for the vibrational modes, that could be expected if there are "helicity-changing" Raman active modes, also known as "chiral" phonons[37–39]. This indicates that the observed Raman phonons in this material are not sensitive to pseudo-angular momentum (PAM) selection rules, despite the presence of I-In-Se helical chains or anisotropy in its crystal structure. Indeed, given the non-chiral space group of the crystal a direct chiral response is not expected. This experimental result verifies that there is no chiral domain in the material despite the presence of helices in the unit cell, and helical chains with both handedness equally coexist as determined in the crystal structure (see Fig. 1b). This is in contrast with α-quartz,[40,41] trigonal Te[24,42,43] or α-HgS[44], whose enantiomeric crystal structures are composed of only right- or left-handed helices. Isolated helices of InSeI may be expected to show the chiral response. Our results suggest that such response to PAM must be associated to a chiral space group, possibly effective with transfer of angular momentum to the lattice vibrations, rather than just to the presence of helical arrangement.[45]

**Conclusions**

We presented a comprehensive and detailed study of the lattice dynamics of an unreported tetragonal InSeI polymorph, confirmed by XRD and STEM investigation, suggesting also new developments in the structural studies of this and similar 1D materials. InSeI was studied here in NWs and bulk crystals by linear and circularly polarized Raman spectroscopy. We demonstrated that angle-dependent linearly polarized Raman spectroscopy measurements in VV (parallel) configuration on the (100)/(010) crystallographic planes allow determining the direction of the helical chains, which align with the crystal growth. In particular, the maximum in the intensity angular dependence of the mode centered at 143 cm$^{-1}$ related to the Se atoms vibrations is aligned to the chain axis. Moreover, we show that it is possible to distinguish between the two types of crystallographic faces accessible, (100)/(010) with the helical chains laying and (001) perpendicular to the chains, thanks to the different pattern of the angular dependence of some of the observed Raman modes. We further analyze the angular dependence of the vibrational modes by Raman tensor analysis leading to the assignment of the representations for each mode. Circularly polarized Raman measurements show that, despite the presence of helical chains and anisotropy in the crystal structure, there are no observable chiral phonons.

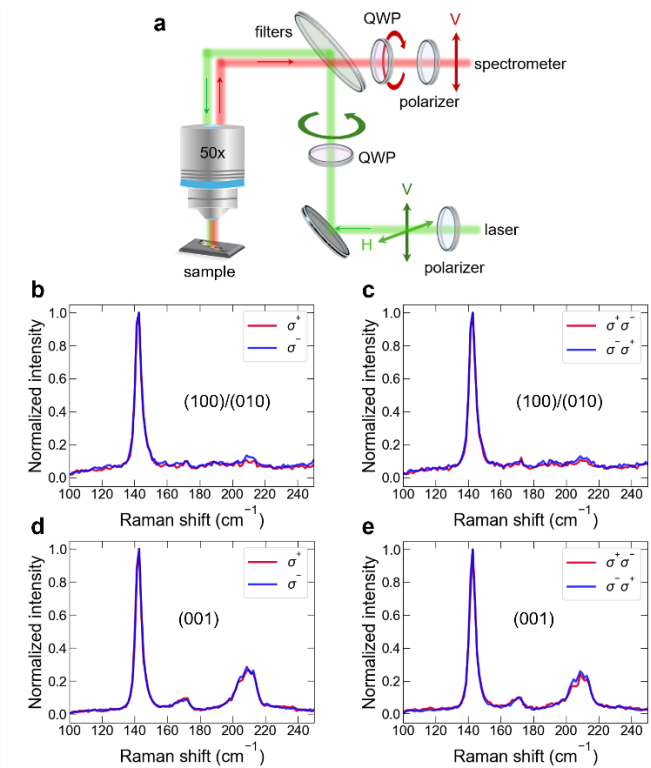

*Fig. 5. Room-temperature circularly polarized and helicity-resolved Raman spectroscopy characterization of bulk crystals. a) Schematic representation of the set-up used to excite the sample and collect the signal with circular polarization placing the corresponding quarter-wave plates (QWP) at the excitation and/or collection paths. b-e) Comparison between the normalized Raman spectra (to the 143 cm$^{-1}$ mode) taken under the configurations $\sigma^+$ vs $\sigma^-$, and $\sigma^+\sigma^-$ vs $\sigma^-\sigma^+$, for incidence on the (100)/(010) planes (b-c), (001) plane (d-e).*

Considering the relevance of knowing the crystal orientation in anisotropic low-dimensional materials to understand in- and out-of-plane directional properties, our results and methodological approach using polarized Raman spectroscopy and tensor analysis, applicable also to other tetragonal materials, are a valuable and non-destructive platform for defining the devices architecture for optical and electrical transport studies.

**Conflicts of interest**

There are no conflicts to declare.

**Data availability**

The data supporting this article have been included as part of the Supplementary Information.

**Supplementary Information (SI)** available: XRD further characterization and crystallographic data; atomic force microscopy characterization of NWs; control experiments; additional polarized Raman spectroscopy study; and Raman tensor analysis. Crystallographic information file for InSeI. See DOI: [10.1039/D4TC04902K](10.1039/D4TC04902K)

**Acknowledgements**

This work is supported under Projects PID2021-122511OB-I00 and PID2021-128004NB-C21 and under the María de Maeztu Units of Excellence Programme (Grant CEX2020-001038-M) funded by Spanish MICIU/AEI/10.13039/501100011033 and by ERDF/EU. Additionally, this work was carried out with support from the Basque Science Foundation for Science (IKERBASQUE), concretely, B.M.G. thanks IKERBASQUE HYMNOS project. L.O.V. thanks the funding from Spanish MICIU/AEI/10.13039/501100011033 and ESF+ for the PhD grant PRE2022-104385. D.S., B.M.-G. and M.G. thanks support from "Ramón y Cajal" Programme by the Spanish MICIU/AEI/10.13039/501100011033 and European Union NextGenerationEU/PRTR (grant nos. RYC2022-037186-I, RYC2021-034836-I and RYC2021-031705-I, respectively). Authors thank R. Llopis (Nanodevices group – CIC nanoGUNE) for his support in the assembly of the Linkam vacuum chamber and Raman instrument optical polarization set-up motorization and implementation; and Dr. E. Goiri Little (Nanodevices group – CIC nanoGUNE) for reading and revising the manuscript.